\documentclass[12pt]{article}

\usepackage{graphicx}
\usepackage{amsmath}
\usepackage{amssymb}
\allowdisplaybreaks

\begin{document}

\baselineskip=17.5pt plus 0.2pt minus 0.1pt

\renewcommand{\theequation}{\arabic{equation}}
\renewcommand{\thefootnote}{\fnsymbol{footnote}}
\makeatletter
\def\CR{\nonumber \\}
\def\pt{\partial}
\def\be{\begin{equation}}
\def\ee{\end{equation}}
\def\bea{\begin{eqnarray}}
\def\eea{\end{eqnarray}}
\def\eq#1{(\ref{#1})}
\def\la{\langle}
\def\ra{\rangle}
\def\hyp{\hbox{-}}


\begin{titlepage}
\title{\hfill\parbox{4cm}{ \normalsize YITP-09-65}\\
\vspace{1cm} Emergent general relativity in the tensor models \\
possessing
Gaussian classical solutions
\footnote{Based on the proceedings of
VIII International Workshop,
``Lie Theory and its Applications in Physics", 
Varna, 15 - 21 June 2009, and of XXV Max Born Symposium, ``The Planck Scale'',
Wroclaw, 29 June - 3 July 2009. 
}}
\author{
Naoki {\sc Sasakura}\thanks{\tt sasakura@yukawa.kyoto-u.ac.jp}
\\[15pt]
{\it Yukawa Institute for Theoretical Physics, Kyoto University,}\\
{\it Kyoto 606-8502, Japan}}
\date{}
\maketitle
\thispagestyle{empty}
\begin{abstract}
\normalsize
This paper gives a summary of the author's works 
concerning the emergent general relativity in a particular class of tensor models,
which possess Gaussian classical solutions.
In general,
a classical solution in a tensor model may be physically regarded as 
a background space, and small fluctuations about the solution as emergent fields on the space.
The numerical analyses of the tensor models possessing Gaussian classical
background solutions
have shown that the low-lying long-wavelength fluctuations around the backgrounds 
are in one-to-one correspondence with the geometric fluctuations on flat spaces 
in the general relativity. 
It has also been shown that part of 
the orthogonal symmetry of the tensor model spontaneously 
broken by the backgrounds can be identified with the local translation
symmetry of the general relativity.
Thus the tensor model provides
an interesting model of simultaneous 
emergence of space, the general relativity, and its local gauge symmetry of translation. 
\end{abstract}
\end{titlepage}

\section{Introduction}
The tensor model was originally considered in 
\cite{Ambjorn:1990ge,Sasakura:1990fs,Godfrey:1990dt}
to generalize the matrix model,
which describes the two-dimensional simplicial quantum gravity, 
to higher dimensional cases.
While the matrix model is a successful tool to analyze the two-dimensional 
simplicial gravity,
the tensor model has not been successful in this direction, 
partly because of the absence of analytical methods to solve it 
and of physically appropriate interpretations of its partition function.

In ref.~\cite{Sasakura:2005js}, 
a new interpretation of the rank-three tensor model was proposed.
Namely,
theory of a dynamical rank-three tensor may be regarded as that of
dynamical fuzzy spaces.
This proposal is based 
on the fact that 
a fuzzy space\footnote{In this paper, 
this terminology is used in its widest meaning. It represents
nonassociative spaces \cite{Sasai:2006ua} 
as well as noncommutative ones \cite{Connes:1994yd}.} 
is described by an algebra of functions,
which can be characterized by a rank-three tensor that defines multiplication, 
$f_a \star f_b=C_{ab}{}^cf_c$. 

This reinterpretation of the tensor model provides a new
practical manner of extracting physics from the tensor model.
In the original interpretation, it is necessary to compute the tensor model 
non-perturbatively, since the large volume limit of spaces corresponds to 
the large loop-number limit of the Feynman diagrams of the tensor model.  
On the contrary, under the new interpretation, the semiclassical treatment of 
the tensor model is
physically meaningful;
its classical solutions
can be regarded as background fuzzy spaces, and small fluctuations around solutions 
as field fluctuations on fuzzy spaces. 
Another key difference from the original proposal
is that rank-three is enough as the rank of tensor to describe fuzzy spaces with 
arbitrary dimensions.
This property drastically simplifies the structures of the tensor model, 
since various dimensional cases can be treated in a common framework. 

The rank-three tensor model has mainly been analyzed in numerical manners
by the present author \cite{Sasakura:2005gv, Sasakura:2006pq, Sasakura:2007sv,
Sasakura:2007ud, Sasakura:2008pe, Sasakura:2009hs}. 
In particular, for the tensor models that possess 
a certain Gaussian type of classical solutions,
it has numerically been shown that the properties of  
low-lying long-wavelength modes of small fluctuations around such Gaussian backgrounds 
are in remarkable agreement with the general relativity in all the dimensional 
cases having been 
studied so far ($D=1,2,3,4$) \cite{Sasakura:2007sv, Sasakura:2007ud, Sasakura:2008pe, 
Sasakura:2009hs}. 
Namely, the general relativity was found to emerge in the tensor model 
as an effective long-wavelength description
of the tensor model around a particular class of classical background solutions. 
This is also expected to be true in any other dimensions, 
since the framework and the procedure of analysis are common.
This paper gives a summary of the results obtained so far 
concerning the emergence of the general relativity in the tensor model.

\section{Tensor models}
There exist various versions of the rank-three tensor model 
\cite{Sasakura:2005js,Sasakura:2005gv}. 
The simplest is the one that has a real symmetric rank-three
tensor as its only dynamical variable, and has the invariance under the orthogonal group.
In this paper, this simplest one is considered.

The dynamical variable is a real-valued rank-three tensor $C_{abc}$, 
each index of which takes integers, $1,2,\ldots, N$.
The number $N$ is the total number of linearly independent 
functions on a fuzzy space, or can more physically be interpreted as 
the number of ``points'' forming a fuzzy space.
The variable $C_{abc}$ is assumed to be totally symmetric,
\begin{equation}
\label{eq:csym}
C_{abc}=C_{bca}=C_{cab}=C_{bac}=C_{acb}=C_{cba}.
\end{equation}
The algebra of products defined by $f_a\star f_b=C_{ab}{}^cf_c$
is commutative but nonassociative in general. Therefore the tensor model 
in this paper  
is a theory of dynamical commutative {\it nonassociative} fuzzy spaces.

The basis of functions $\{ f_1,\,f_2,\ldots,f_N \}$ can be changed by 
linear transformations.
A simple choice of equivalence of the basis functions is to assume that 
bases related by the orthogonal transformations represent equivalent fuzzy spaces.
Correspondingly, the tensor model must be invariant under the orthogonal transformation
\begin{equation}
\label{eq:trans}
C_{abc}\rightarrow M_a{}^{a'}   M_b{}^{b'} M_c{}^{c'} C_{a'b'c'},
\end{equation}
where $M_a{}^{a'}$ is an arbitrary element of the orthogonal group $O(N)$.

The definition of the system is given by a partition function,
\begin{equation}
\label{eq:partition}
Z=\int \prod dC_{abc}\ e^{-S(C)},
\end{equation} 
where $S(C)$ is an action with the variable $C_{abc}$, and must be invariant under the 
orthogonal group transformation (\ref{eq:trans}). 
The integration measure $\prod dC_{abc}$ must also be invariant under (\ref{eq:trans}),
and is defined from the invariant metric in the space of $C_{abc}$ given by 
\begin{equation}
\label{eq:Cmeasure}
ds_C^2=d C_{abc}\, d C_{abc}.
\end{equation}

\section{Gaussian backgrounds}
\label{sec:gauss}
So far, the emergence of the general relativity in the tensor model has only been  
shown around a particular class of backgrounds in the tensor model. 
These backgrounds have certain Gaussian forms, and the algebras defined
by them represent certain simple kinds of commutative {\it nonassociative}  fuzzy flat 
spaces with arbitrary dimensions \cite{Sasai:2006ua}.
In this section, to describe such genuine Gaussian backgrounds,
the indices of $C_{abc}$ are assumed to  take continuous values,
while they will take finite discrete values in the actual analyses of the
following sections.
 
The Gaussian backgrounds have the form,   
\begin{equation}
\bar C_{x_1x_2x_3}=B \exp\left[ -\beta \left( (x_1-x_2)^2+(x_2-x_3)^2+(x_3-x_1)^2\right)\right],
\label{eq:cx}
\end{equation}
where $B$ and $\beta$ are positive numerical constants, 
$x_i$ are $D$-dimensional continuous coordinates,
$x_i=(x_i^1,x_i^2,\cdots,x_i^D)$, and  $(x)^2=\sum_{\mu=1}^D (x^\mu)^2$.  
The algebra of functions $f_{x_1}\star f_{x_2}=\bar C_{x_1x_2}{}^{x_3}f_{x_3}$ defines
a commutative {\it nonassociative} fuzzy $D$-dimensional 
flat space considered in \cite{Sasai:2006ua}.
Because of the translational symmetry of (\ref{eq:cx}), it is generally 
more convenient to describe it in the momentum basis. 
By applying Fourier transformation to the coordinate indices, 
one obtains the expression in the momentum basis\footnote{
Generally in the momentum basis, the tensor must satisfy 
$C_{-p_1,-p_2,-p_3}^*=C_{p_1p_2p_3}$ because of the reality condition of the tensor
in the coordinate basis, and one needs also an 
additional symmetric tensor $g^{p_1p_2}=\delta(p_1+p_2)$ 
for contracting the indices. These details are essentially important in the mode 
analysis of the following sections.} as  
\begin{equation}
\label{eq:cp}
\bar C_{p_1p_2p_3}=A \,
\delta^D(p_1+p_2+p_3) \exp \left[ -\alpha \left( p_1^2+p_2^2+p_3^2\right) \right], 
\end{equation}
where $A$ and $\alpha= 1/(12 \beta)$ are positive numerical constants.

One of the motivations for considering such particular solutions is the (partial) 
computability due to the Gaussian forms. 
Another is that the fuzzy spaces are physically well-behaved, because
the fuzziness is well localized and the spaces are invariant under 
the Poincare transformation, representing fuzzy $D$-dimensional flat spaces. 
Moreover, as shown in the next section, there exists 
a natural correspondence between the metric field in the general relativity 
and the tensor around the Gaussian backgrounds in the tensor model.

\begin{figure}
\centerline{\includegraphics[scale=.8]{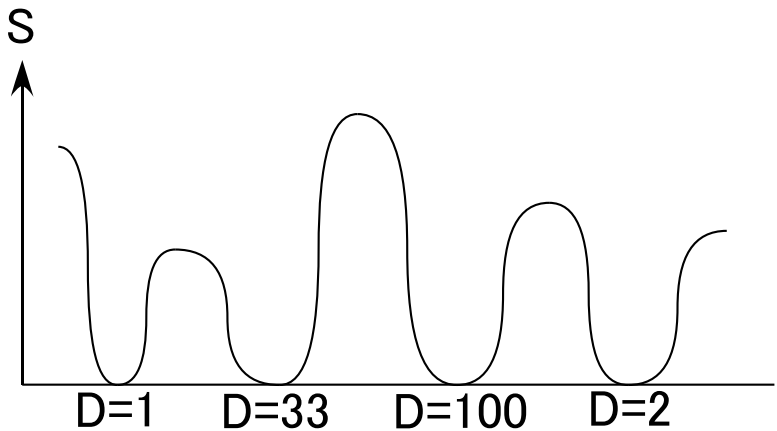}}
\caption{Illustration of an action possessing Gaussian backgrounds as its classical solutions.}\label{fig:potential}
\end{figure}

In fact, there exist infinitely many actions
that have such Gaussian backgrounds as their
classical solutions. 
Two explicit examples have been studied so far 
\cite{Sasakura:2007sv, Sasakura:2007ud, Sasakura:2008pe, Sasakura:2009hs}.
Unfortunately, the explicit forms of the actions are very complicated and unnatural. 
This is a serious problem, which must be investigated in future study.
However, what is interesting and remarkable in these actions in common
is that each of them
contains all the dimensional Gaussian fuzzy flat spaces as its classical 
solutions, as illustrated in Figure~\ref{fig:potential}.
This means that all the dimensional spaces can be treated with one action 
in a unified manner. Thus, for example, it is in principle possible to 
study transitions between spaces with distinct dimensions in the tensor model.

\section{Correspondence between general relativity and tensor models around
Gaussian backgrounds }
In the physical interpretation of the tensor model, a classical solution should
be regarded as a background space, and fluctuations of 
tensor around such a classical solution as field fluctuations on a background
space. The main interest of the present study is whether such fluctuations can be
identified with the general relativity or not.
To check this, the procedure carried out so far 
in \cite{Sasakura:2007sv, Sasakura:2007ud, Sasakura:2008pe, Sasakura:2009hs} assumes 
a correspondence between the tensor model around the Gaussian backgrounds and 
the metric tensor field in the general relativity. 
This correspondence enables one to compute the expectations about the tensor model 
from the general relativity. 
If the expectations successfully agree with the numerical analysis of the tensor model, 
one may conclude that 
the general relativity is emergent around the Gaussian backgrounds in the tensor model.

Generalizing the Gaussian backgrounds (\ref{eq:cx}) in a coordinate invariant manner,
one can derive a natural correspondence between the  
metric tensor field in the general relativity
and the tensor around the Gaussian backgrounds as
\begin{eqnarray}
\label{eq:correspondence}
&&C_{x_1x_2x_3}=B g(x_1)^\frac14 g(x_2)^\frac14 g(x_3)^\frac14 
\exp\left[ - \beta \left( d(x_1,x_2)^2+d(x_2,x_3)^2+d(x_3,x_1)^2\right)\right],
\end{eqnarray}
where $g(x)={\rm det}\left(g_{\mu\nu}(x)\right)$, and $d(x,y)$ denotes
the geometric distance between $x$ and $y$. 
The main assumption in this correspondence
is that the low-lying long-wavelength fluctuations of the tensor 
around the Gaussian backgrounds in the tensor model are exhausted by 
the metric field in the general relativity in the manner 
given in (\ref{eq:correspondence}). It should be noted that 
this correspondence could be modified in higher orders of the fuzziness 
$\alpha\sim 1/\beta$. For example, there could exist corrections such as 
$g_{\mu\nu}(x)\rightarrow g_{\mu\nu}(x)+const.~\alpha R_{\mu\nu}(x)+\cdots$
in (\ref{eq:correspondence}).

Although it is certainly possible to directly use the correspondence 
(\ref{eq:correspondence}) in the comparison between the tensor model and the general 
relativity, it is much more convenient to use a tensor
with a smaller rank. Let me define 
\begin{equation}
K_{ab}=C_{acd}C_b{}^{cd}.
\end{equation}
Small fluctuations $\delta C_{abc}$ around a classical solution $C^0_{abc}$ induces 
fluctuations of $K_{ab}$ as 
\begin{equation}
\label{eq:tensordk}
\delta K_{ab}=\delta C_{acd}\, {C^0{}_b}^{cd}+  C^0_{acd}\, {\delta C_b}^{cd}.
\end{equation}

On the other hand, if one assumes a Gaussian background and 
puts the assumed correspondence (\ref{eq:correspondence}) into 
(\ref{eq:tensordk}), one obtains in the lowest order \cite{Sasakura:2007ud} 
\begin{equation}
\label{eq:metricdk}
\delta K_{p_1p_2}
=
\delta g_{\mu\nu}(p_1+p_2)\, 
(p_1-p_2)^\mu (p_1-p_2)^\nu \exp\left(-\frac1{16\beta}(p_1-p_2)^2 \right), 
\end{equation}
where the momentum basis is used, 
and $\delta g_{\mu\nu}(p)$ is the Fourier transform of $\delta g_{\mu\nu}(x)$,
which describes the fluctuations of the metric tensor field around a flat background.

In later sections, the analysis of eigenvectors
gives $\delta C_{abc}$ of each fluctuation mode in the tensor model. 
Then one can compute $\delta K_{ab}$ of each mode by putting this $\delta C_{abc}$
and the background $C^0_{abc}$ into (\ref{eq:tensordk}). This $\delta K_{ab}$ obtained
from the numerical analysis of the tensor
model can be compared with the fluctuation modes of the metric tensor
field in the general relativity through (\ref{eq:metricdk}).

Another important fact which can be derived from (\ref{eq:correspondence}) is that 
the measure which must be used in the analysis of the general relativity is 
uniquely determined from the measure (\ref{eq:Cmeasure}) in the tensor model. 
By putting the correspondence (\ref{eq:correspondence}) into (\ref{eq:Cmeasure}), 
one obtains
the DeWitt supermetric \cite{DeWitt:1962ud},
\begin{equation}
\label{eq:supermetric}
ds^2_{DW}=
 \int d^Dx \sqrt{g(x)} \left[
 \left(g^{\mu\nu}(x)\delta g_{\mu\nu}(x)\right)^2
+4 g^{\mu\rho}(x)g^{\nu\sigma}(x) \delta g_{\mu\nu}(x) 
\delta g_{\rho\sigma}(x)\right], 
\end{equation}
in the lowest order \cite{Sasakura:2007ud}.

\section{Geometric fluctuations on flat tori}
\label{sec:gr}
In this section, I will study the small geometric fluctuations on
$D$-dimensional flat tori in the general relativity to prepare  
for the comparison with the tensor model.
The important point in the analysis is that not all of the 
fluctuations of the metric tensor field are the fluctuations of 
geometry, because of the gauge symmetry (local translation
symmetry) in the general relativity.
The relevant modes are only 
those which are normal to the gauge symmetry \cite{Sasakura:2007ud, Sasakura:2008pe}.
The measure to be used to define this normality condition
is the DeWitt supermetric given in (\ref{eq:supermetric}),
since the numerical analysis of the tensor model uses the corresponding
measure (\ref{eq:Cmeasure}) as shown in the following section. 
 
For small fluctuations around a flat metric 
$g_{\mu\nu}=\delta_{\mu\nu}+\delta g_{\mu\nu}$,
the supermetric (\ref{eq:supermetric}) is given by
\begin{equation}
\label{eq:explicit}
ds^2_{DW}=
\int d^D x \Bigg[
\sum_{\mu=1}^D
5 (\delta g_{\mu\mu})^2+
\sum_{{\mu=1,\nu=2,}\atop{\mu<\nu}}^D \left\{ 2 \delta g_{\mu\mu}\delta g_{\nu\nu}+ 
8 (\delta g_{\mu\nu})^2\right\}
\Bigg].  
\end{equation}
On the other hand,
the infinitesimal gauge transformation on a flat background is given 
in the momentum basis by
\begin{equation}
\label{eq:gaugemom}
\delta_v g_{\mu\nu}(p)=i p_\mu v_\nu(p)+i p_\nu v_\mu(p),
\end{equation}
where $v_\mu(p)$ is the Fourier transform of local translation vector.

At the vanishing momentum sector $p_\mu=0$,
as can be seen in (\ref{eq:gaugemom}), 
the gauge transformation is vacant, 
and all the components of the metric tensor are geometric degrees of freedom.
By diagonalizing the supermetric (\ref{eq:explicit}), 
the modes can be shown to be classified into the following three orthogonal classes:\\
(i) 1 conformal mode: $\delta g_{\mu\mu}=\delta g_{\nu\nu}$.\\
(ii) $D-1$ traceless diagonal modes: $\sum_{\mu=1}^D \delta g_{\mu\mu}=0$.\\
(iii) $D(D-1)/2$ off-diagonal modes.

At the nonvanishing momentum sector, one may take the momentum to be in the direction
$(p_1,0,\cdots,0)$ with obvious generalization to the other directions.
Then one can show that 
the modes normal to the gauge directions (\ref{eq:gaugemom}) 
in the sense of the supermetric (\ref{eq:explicit}) can
be classified into the following three orthogonal classes:\\
(i) 1 diagonal mode: $\delta g_{11}=-D+1$, $\delta g_{\mu\mu}=5$ for $\mu\neq 1$. \\
(ii) $D-2$ traceless diagonal modes: $\delta g_{11}=0$, $\sum_{\mu=2}^D\delta g_{\mu\mu}=0$.\\
(iii) $(D-1)(D-2)/2$ off-diagonal modes: $\delta g_{\mu\nu}\neq 0$ for $\mu\neq \nu$, $\mu,\nu \neq 1$. 

$\delta K_{ab}$ for the geometric fluctuations of each mode in the general relativity
can be computed by putting these results into (\ref{eq:metricdk}), and 
can be compared with the numerical analysis of the tensor model.

\section{Numerical analysis of small fluctuations around Gaussian backgrounds in tensor models}
In this section, I will give a brief summary of the numerical analyses
having been done so far on the small fluctuations around the Gaussian backgrounds in the tensor model
\cite{Sasakura:2007sv, Sasakura:2007ud, Sasakura:2008pe}. 

To unambiguously determine the spectra, the fluctuations must
be normalized in a manner respecting the symmetries of the tensor model. 
The tensor $C_{abc}$ is symmetric under the exchanges of the indices as 
in (\ref{eq:csym}), and the tensor model 
has the  orthogonal group symmetry (\ref{eq:trans}).
Thus the independent components of fluctuations are 
normalized through (\ref{eq:Cmeasure}) as
\begin{equation}
\label{eq:Cmeasuresym}
ds^2_C=\sum_{(abc)} m[(abc)]\, 
d C_{abc}\, d C^{abc}=\sum_{(abc)} d \tilde C_{(abc)}
\, d \tilde C^{(abc)},
\end{equation}
where $(abc)$ denotes an order-independent set of three indices, 
$\tilde C_{(abc)}\equiv\sqrt{m[(abc)]}\,C_{abc}$ are the independent
normalized components, and $m[(abc)]$ is the multiplicity defined by
\begin{equation}
m[(abc)]=\left\{ 
\begin{array}{ll}
1 & \hbox{for } a=b=c,\\
3 & \hbox{for } a=b\neq c,\ b=c\neq a,\ c=a\neq b,\\
6 & \hbox{all different}.
\end{array}
\right.
\end{equation}
Then the coefficient matrix for
the normalized fluctuations in the quadratic order
around a background $C^0$ is given by
\begin{equation}
\label{eq:M}
M^{(abc),(def)}\equiv \frac{1}{2} 
\frac{\partial^2 S}{\partial \tilde C_{(abc)}\partial \tilde C_{(def)}}
\Bigg|_{C=C^0}.
\end{equation}

The numerical analysis is carried out
to obtain the eigenvalues and eigenvectors of the matrix 
(\ref{eq:M}), and then the results are
compared with the general relativity through the 
procedure explained in the previous sections.
The genuine Gaussian backgrounds presented in Section \ref{sec:gauss}
cannot be used as a background $C^0$, 
because of their infinite number of degrees of freedom.
Thus the actual numerical analyses have been done around 
the backgrounds of $D$-dimensional fuzzy flat tori.
To describe such backgrounds of fuzzy flat tori in terms of $C_{abc}$, 
the indices are assumed to take integer momenta bounded by a cut off,
and the momentum conservation, $C^0_{p_1p_2p_3}\propto 
\delta_{p_1+p_2+p_3,\vec 0}$, is assumed on account of 
the translational symmetry of such flat tori. 
In fact, under these assumptions, 
one can numerically find classical solutions
that resemble the genuine Gaussian 
backgrounds \cite{Sasakura:2007sv,Sasakura:2007ud}. 
Using these numerical classical solutions as backgrounds,
the coefficient matrices (\ref{eq:M}) have been analyzed for dimensions $D=1,2$
in \cite{Sasakura:2007sv,Sasakura:2007ud}, and also for $D=1,2,3,4$ with an approximate
method explained below \cite{Sasakura:2007sv,Sasakura:2008pe}.

The numerical analyses have shown that the spectra of the modes with long-wavelengths 
can roughly be classified into the following three classes.\\
(i) The ``heavy'' modes with spectra of order 1 or larger.\\
(ii) In $D>1$, there exist low-lying modes 
with non-vanishing spectra of order much smaller than 1.\\
(iii) The zero modes with vanishing spectra.\\
It is observed that the two classes (i) and (ii) are rather clearly separated so that
they are located hierarchically.

The spectra in the class (ii) have been shown to form trajectories of the fourth 
power of momenta. Moreover, the spectral patterns and the mode profiles have been shown
to be in good agreement with the geometric degrees of freedom in the general relativity
discussed in Section \ref{sec:gr},
by comparing (\ref{eq:tensordk}) computed numerically from the
tensor model and (\ref{eq:metricdk}) from the general relativity. 
This shows that
the general relativity is emergent around such backgrounds, and that 
the lowest effective actions are composed of curvature quadratic terms.

As for the class (iii),
by counting their numbers, these modes have been identified 
with the modes of the $O(N)$ symmetry transformations in the tensor model
spontaneously broken to the remaining symmetry $SO(2)^D$ of the torus backgrounds. 
Therefore these zero spectra are just the gauge modes of the tensor model.
The relation between these zero modes and the local 
gauge symmetry (local translation symmetry) of the general relativity 
has been studied in \cite{Sasakura:2009hs}\footnote{
In fact, the idea to regard gauge symmetries as spontaneously broken symmetries 
is rather old. For example, see \cite{Ferrari:1971at,Brandt:1974jw,Borisov:1974bn}.}, 
which will be summarized in the following section.

\begin{figure} 
\centerline{\includegraphics[scale=.5]{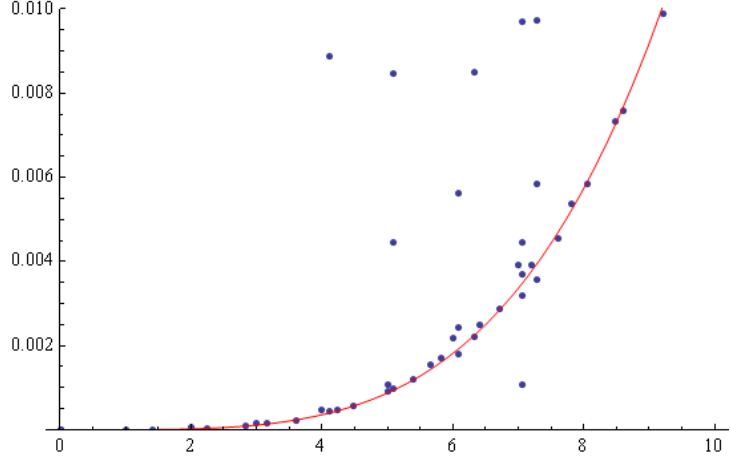}}
\caption{The low-lying spectra around the background of
a $D=2$ fuzzy flat torus for cut-off 
$L=10$ and $\alpha=1.5/L^2$. The horizontal axis is the size of momentum 
$|p|=\sqrt{(p^1)^2+(p^2)^2}$ of fluctuation modes, and the vertical axis
is the spectral value. The solid line is $1.4\times 10^{-6} |p|^4$.}
\label{fig:dim2L10spec}
\end{figure}

\begin{figure} 
\centerline{
\includegraphics[width=6cm]{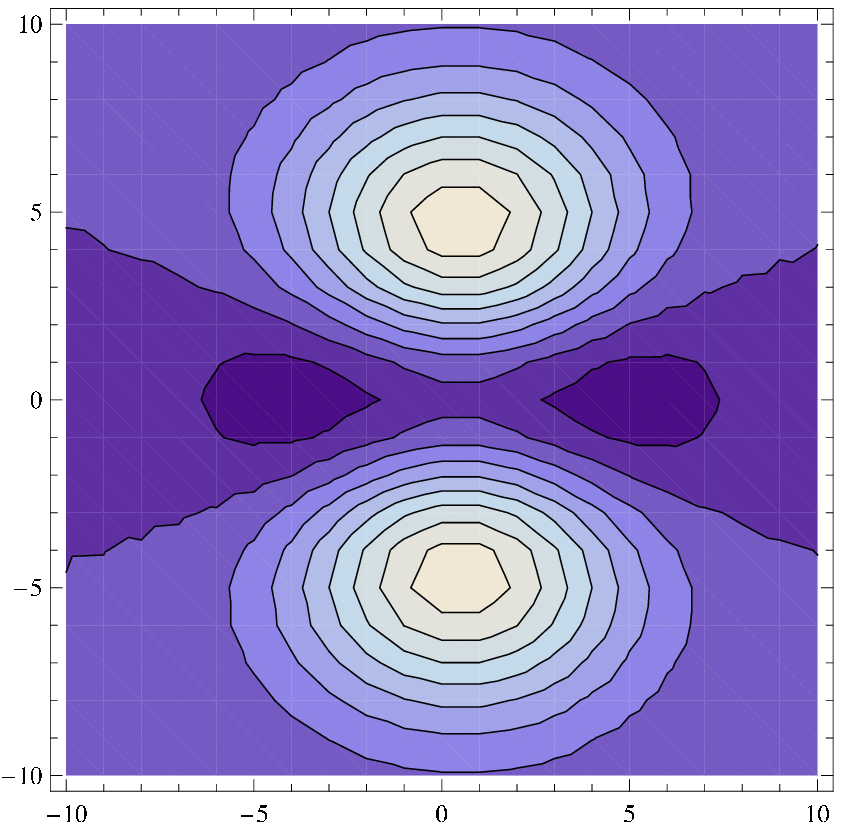}
\hspace{2cm}
\includegraphics[width=6cm]{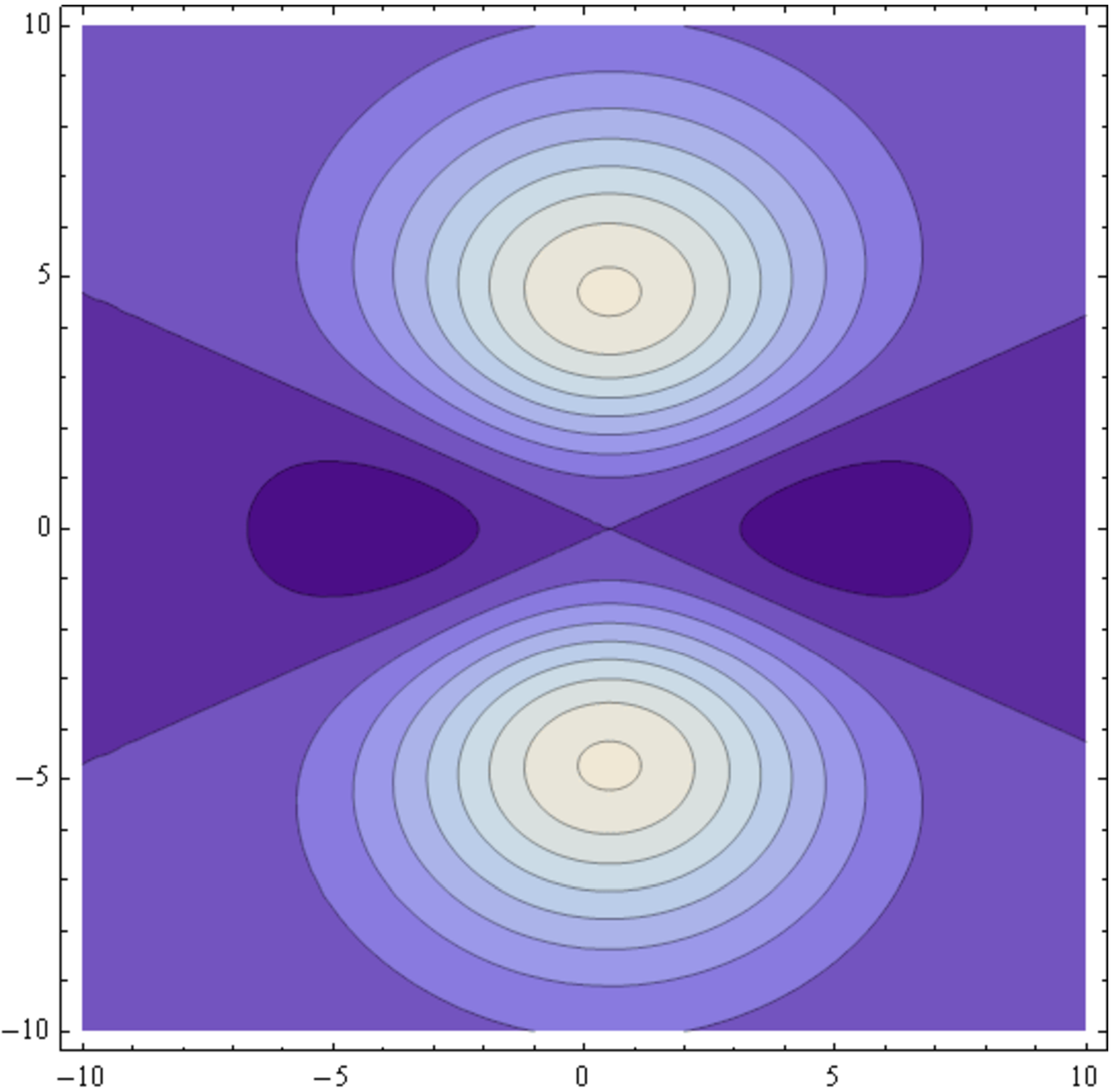}
}
\caption{The contour plots of $\delta K_{q,-q+p}$ 
for the low-lying mode at $p=(1,0)$ sector for $D=2$.
The left and right figures are (\ref{eq:tensordk}) from the numerical analysis of 
the tensor model
and (\ref{eq:metricdk}) for the diagonal mode in the general relativity, respectively.
The axes are $q=(q^1,q^2)$.}
\label{fig:dim2L10}
\end{figure}

In the approximate method used in \cite{Sasakura:2007sv,Sasakura:2008pe}, the backgrounds $C^0$ in (\ref{eq:M}) are not taken to be the
classical solutions, but to be approximate ones, the Gaussian backgrounds (\ref{eq:cp})
with the modifications that 
the continuum $\delta$-functions are replaced with Kronecker deltas, 
and that $\alpha$ is taken in the range $1/L^2 \lesssim \alpha \lesssim 1/L$
for the approximation to be good.
In this approximation, it is easier to numerically analyze the cases with 
larger $L$, and 
the agreement between the spectra in the class (ii) and the geometric degrees of 
freedom in the general relativity has been observed more clearly than 
without approximations.
Part of the results for $D=2$ are shown 
in Figures \ref{fig:dim2L10spec} and \ref{fig:dim2L10}. 
Especially, Figure \ref{fig:dim2L10} shows very clearly the agreement of 
the mode profiles 
between the tensor model and the general relativity.

\section{BRST gauge fixing of tensor models and emergent ghost fields}
The connection between the spontaneously broken $O(N)$ symmetry and the 
local gauge symmetry (local translation symmetry) in the general relativity 
has been discussed in \cite{Sasakura:2009hs}.
In the paper, the BRST gauge fixing procedure has been applied to the
$SO(N)$ symmetry of the tensor model,
and the spectra of the ghost quadratic term have been studied numerically.
Then the appearance of emergent massless ghost fields has been observed, 
and they have been
identified with the reparametrization ghost fields coming from the BRST 
gauge fixing of the general relativity.

Let me start with the gauge fixing in the tensor model.
The off-shell nilpotent BRST transformation in the tensor model is defined by
\begin{equation}
(\delta_B C)_{abc}=c_i(T^i C)_{abc},\ 
\delta_B c_k=\frac12 f^{ij}{}_k c_i c_j,\ 
\delta_B \bar{c}_j=i B_j, \ 
\delta_B B_i=0.
\end{equation}
where $c_i(\bar c_i)$ and $B_i$ are the ghosts (anti-ghosts), and the bosonic 
auxiliary variables \cite{Kugo:1981hm}, respectively.
Here $(T^i C)_{abc}$ is the infinitesimal orthogonal transformation defined by
\begin{equation}
(T^i C)_{abc}
\equiv T^i{}_{a}{}^{a'} C_{a'bc}+T^i{}_b{}^{b'}C_{ab'c}+T^i{}_c{}^{c'}C_{abc'},
\end{equation}
where $T^i$ are the elements of the Lie algebra $so(N)$ in the vector representation, and
$f^{ij}{}_k$ is the structure constant defined by $[T^i,T^j]=f^{ij}{}_k T^k$.

\begin{figure}
\centerline{
\includegraphics[width=8cm]{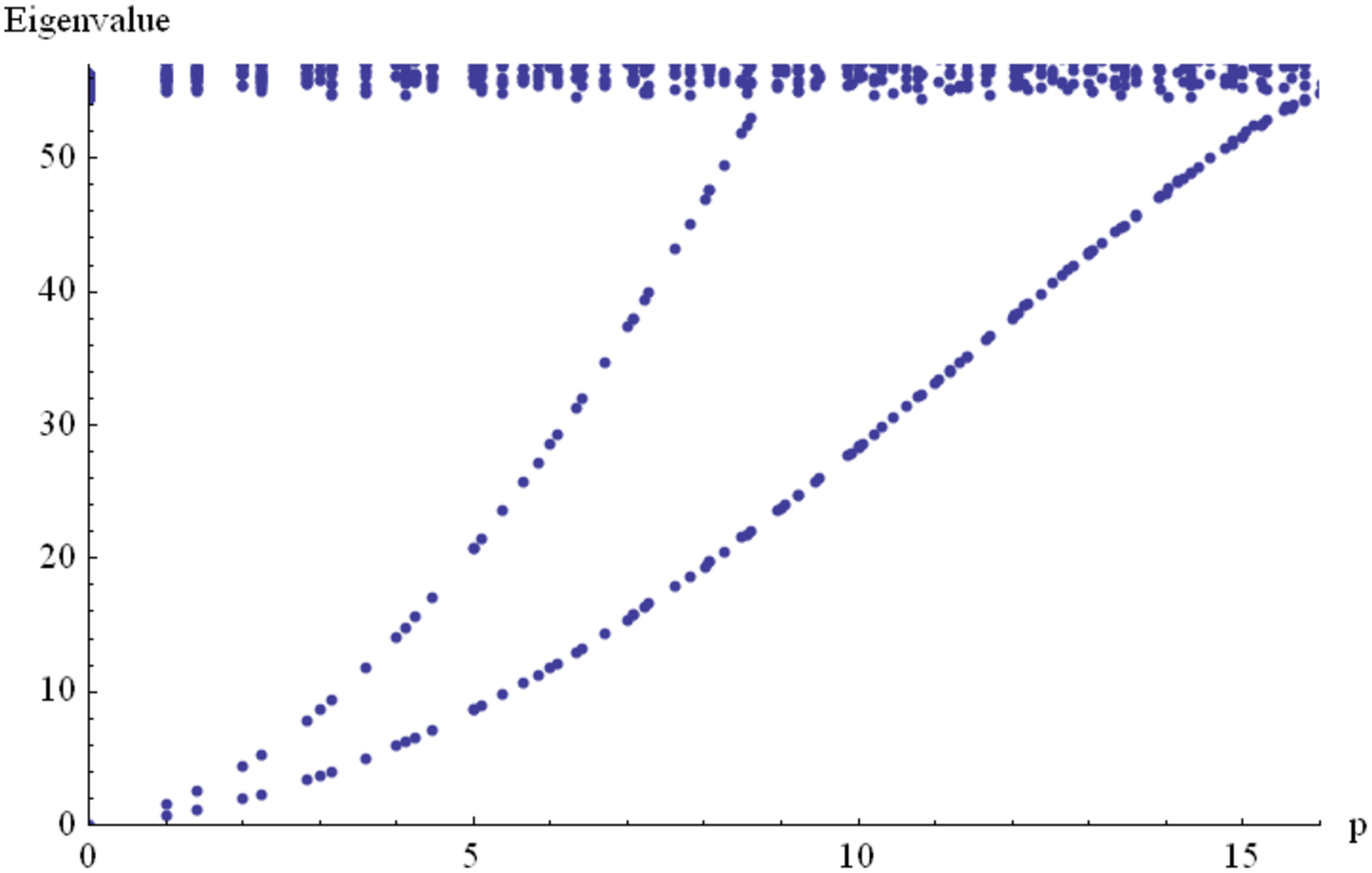}
\hspace{.5cm}
\includegraphics[width=8cm]{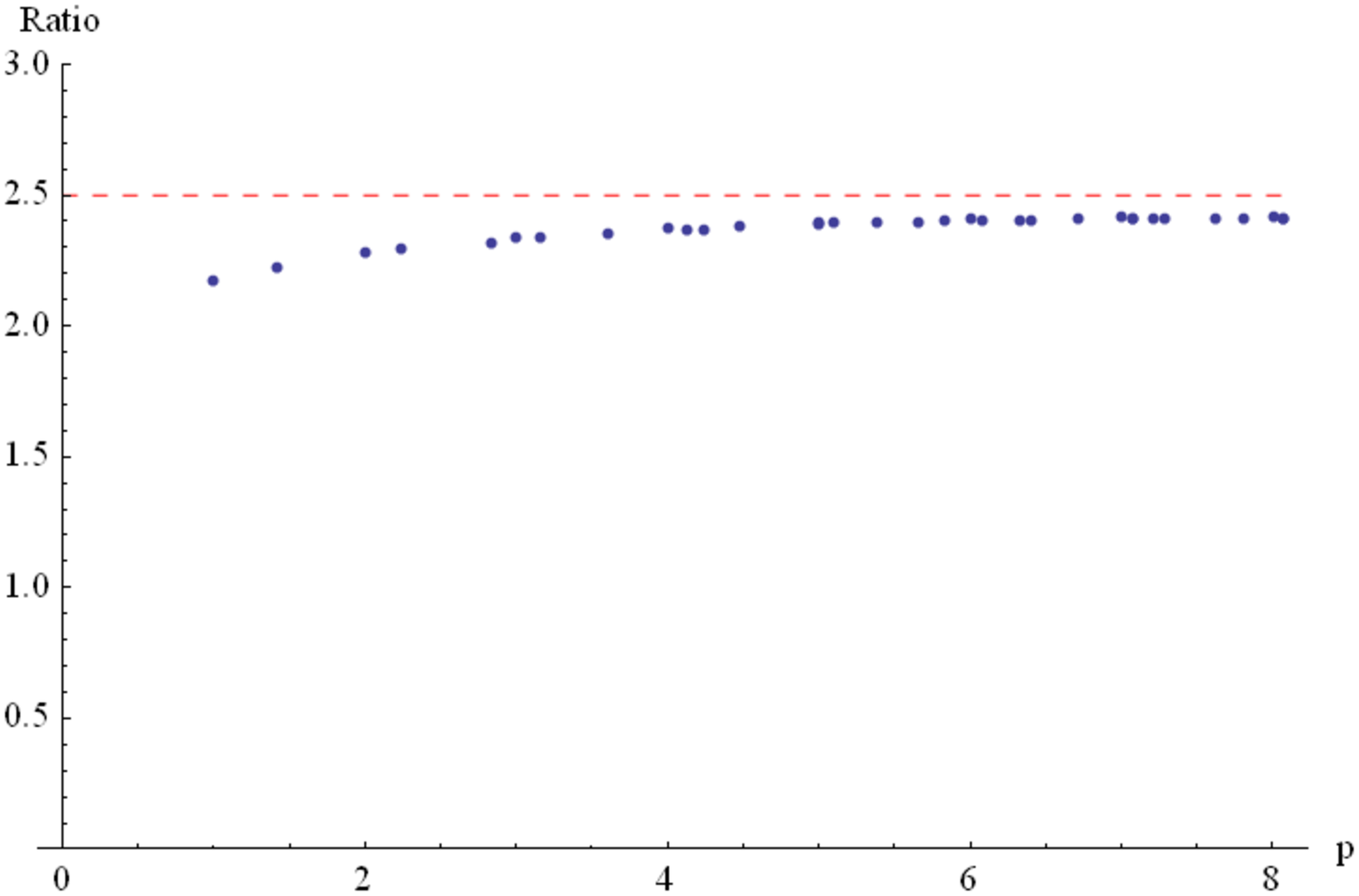}
}
\caption{The left figure shows the low part of the spectra for $D=2$, $L=100$, $\alpha=2/L^2$.
The right figure shows the ratios of the two trajectories.
The horizontal axis is the momentum size $\sqrt{(p^1)^2+(p^2)^2}$.}
\label{fig:ghostspec}
\end{figure}

Now let me define a new dynamical variable $A$ by shifting the original variable 
$C$ by a background $C^0$, 
\begin{equation}
\label{eq:shiftca}
C_{abc}=C^0_{abc}+A_{abc}.
\end{equation}
Then a natural gauge fixing plus Faddeev-Popov action is given by
\begin{equation}
\label{eq:sgftensor}
S_{GF+FP}=\delta_B(\bar c_i F^i), \ 
F^i= \langle T^iC^0,A\rangle, 
\end{equation}
where $\langle \cdot,\cdot \rangle$ is the inner product 
associated with the measure (\ref{eq:Cmeasure}). The reason why this is natural 
is that the gauge fixing conditions ($F^i=0$) only allow $A$ to be normal to the symmetry
directions around the background $C^0$. This action contains the ghost quadratic term as
\begin{equation}
S_{GF+FP}=-\bar c_i \langle T^i C^0, T^j C^0\rangle c_j+\cdots.
\end{equation}
Figure \ref{fig:ghostspec} shows the spectra of this ghost quadratic term and the 
ratio of the two trajectories obtained numerically for a Gaussian background of
$D=2$ flat torus.  

On the other hand, the off-shell nilpotent BRST gauge transformation 
in the general relativity is given by
\begin{equation}
\delta_B g_{\mu\nu}=\nabla_\mu c_\nu+\nabla_\nu c_\mu, \ 
\delta_B c_\mu=-c^\nu \nabla_\mu c_\nu, \ 
\delta_B \bar c_\mu=i B_\mu, \ 
\delta_B B_\mu=0.
\end{equation}
Corresponding to (\ref{eq:shiftca}), let me define
a new dynamical field $h_{\mu\nu}$ by shifting the metric by a flat
background, $g_{\mu\nu}=\delta_{\mu\nu}+h_{\mu\nu}$.
Then, an action naturally corresponding to (\ref{eq:sgftensor}) is given by
\begin{equation}
\label{eq:sgffpingen}
S^{GR}_{GF+FP}=\delta_B \langle \partial_\mu \bar c_\nu+\partial_\nu \bar c_\nu, h_{\rho\sigma}\rangle_{DW},
\end{equation} 
where $\langle \cdot , \cdot \rangle_{DW}$ is the inner product associated with
the DeWitt supermetric (\ref{eq:supermetric}).
In fact, the gauge fixing conditions in \eq{eq:sgffpingen} only allows $h_{\mu\nu}$ 
to be normal to the infinitesimal gauge transformations of the flat background.  
This action contains the ghost kinetic term as 
\begin{equation}
S^{GR}_{GF+FP}= 2 \int d^Dx\left[  
-6\, \partial^\mu \bar c_\mu \, \partial^\nu c_\nu-4\, \partial_\mu \bar c_\nu\, \partial^\mu c^\nu
\right]+\cdots.
\end{equation}
This kinetic term contains 
the longitudinal ($n_\mu \propto p_\mu$) and normal ($n_\mu p^\mu=0$) modes
with spectra $20p^2$ and $8p^2$, respectively.
In fact, 
the ratio of spectra $20/8=2.5$ agrees well with the numerical analysis 
of the tensor model as plotted in Figure \ref{fig:ghostspec}.

\begin{figure}
\centerline{
\includegraphics[width=3.9cm]{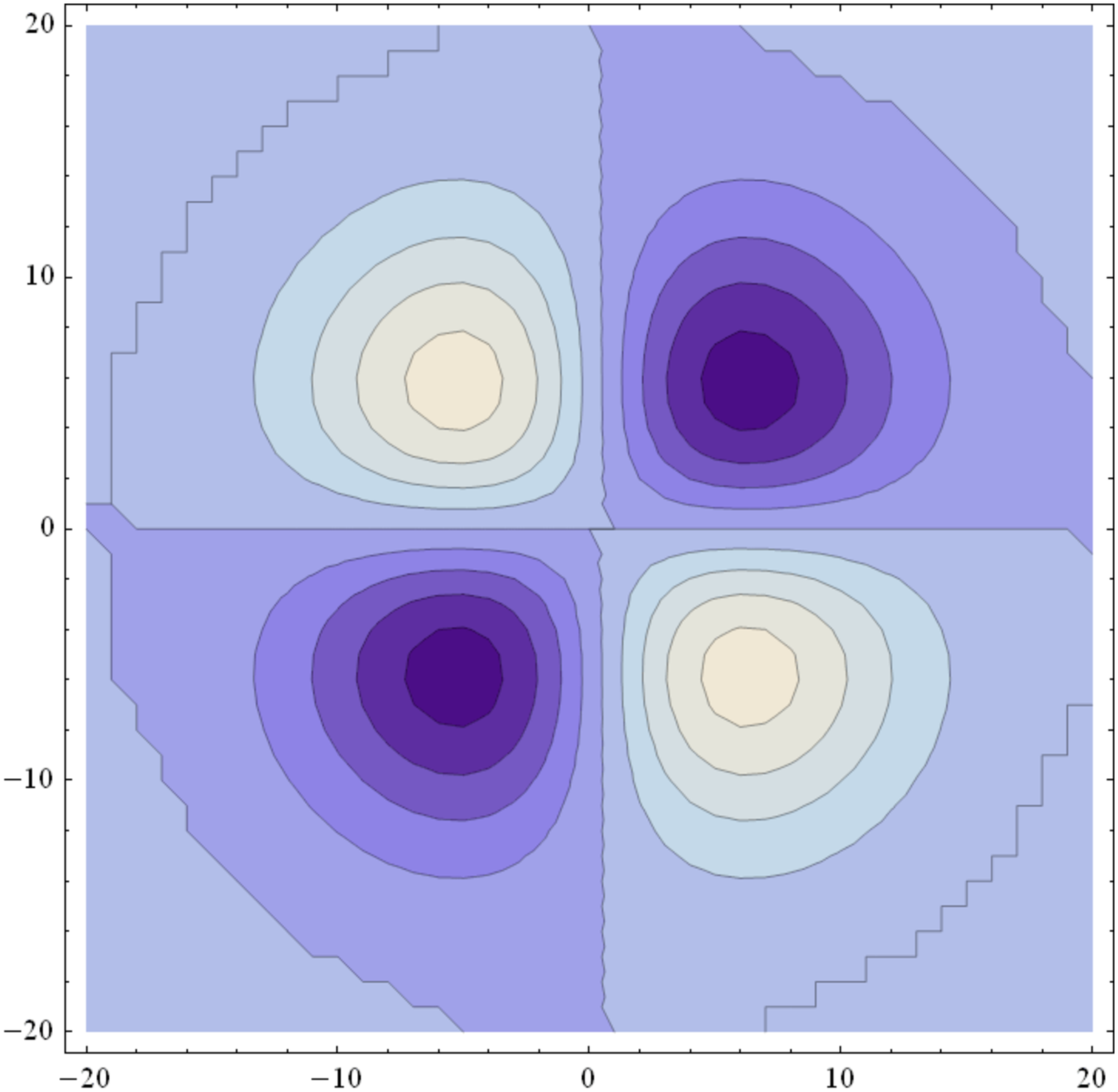}
\includegraphics[width=3.9cm]{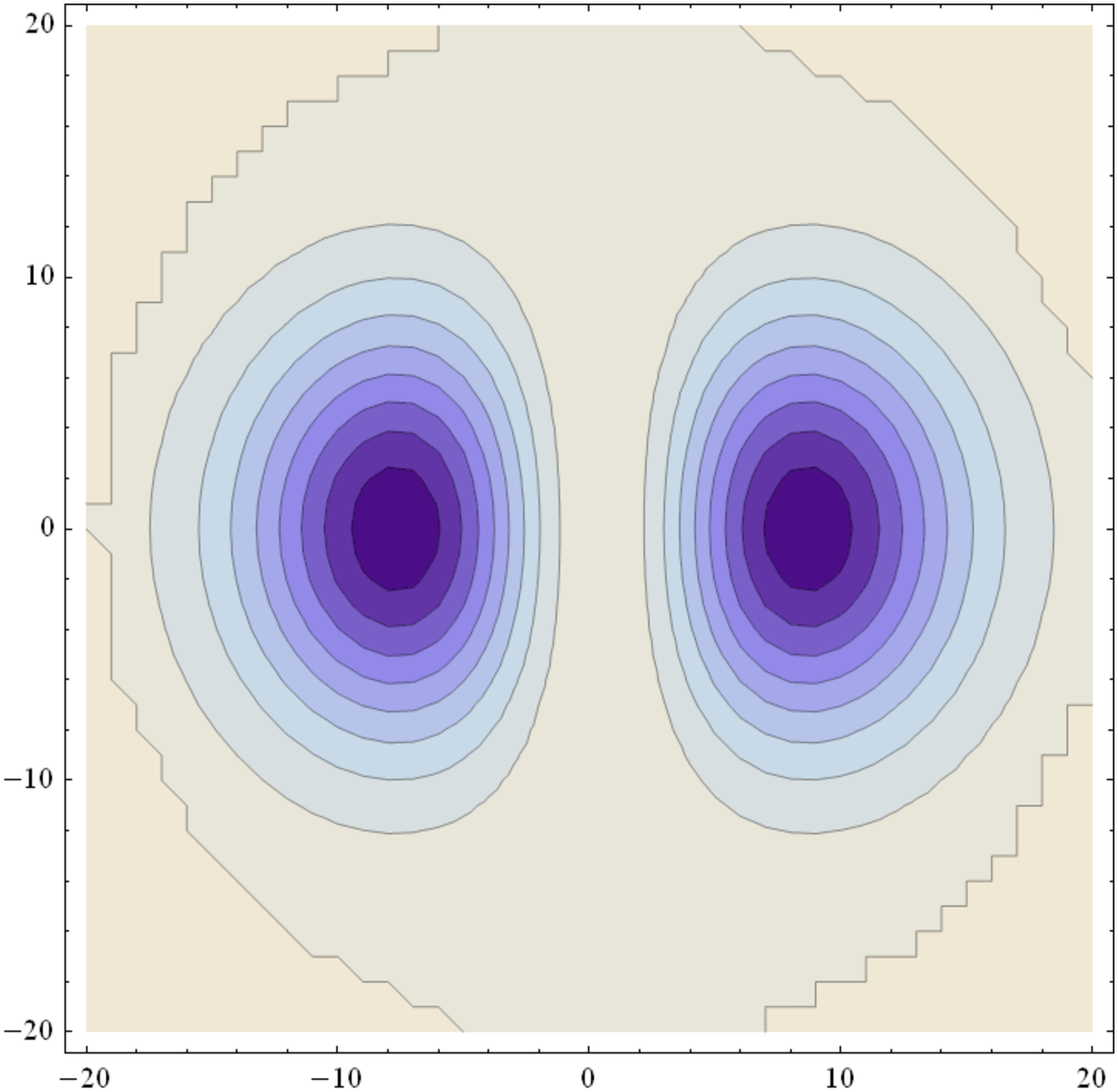}
\hspace{.5cm}
\includegraphics[width=3.9cm]{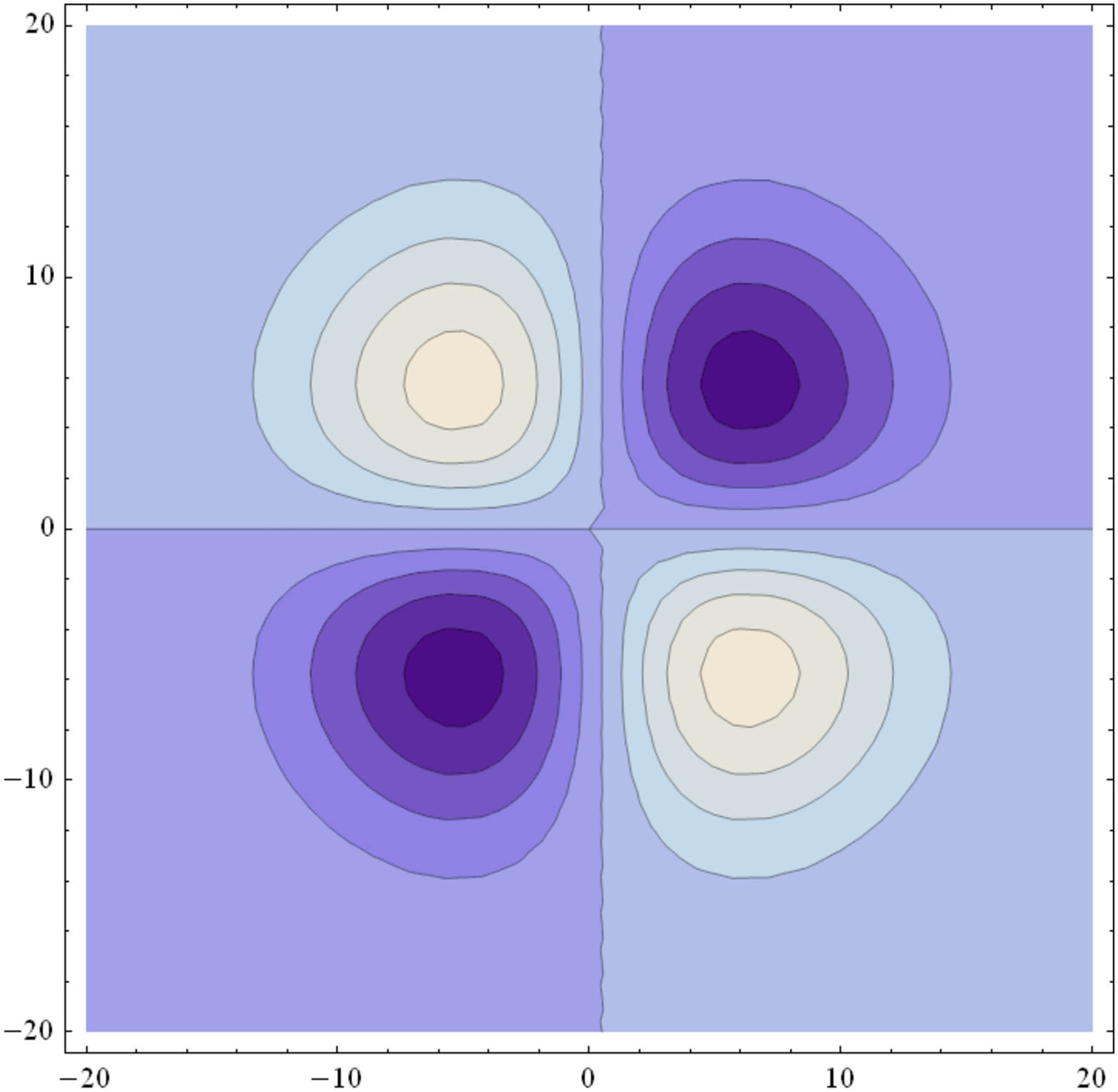}
\includegraphics[width=3.9cm]{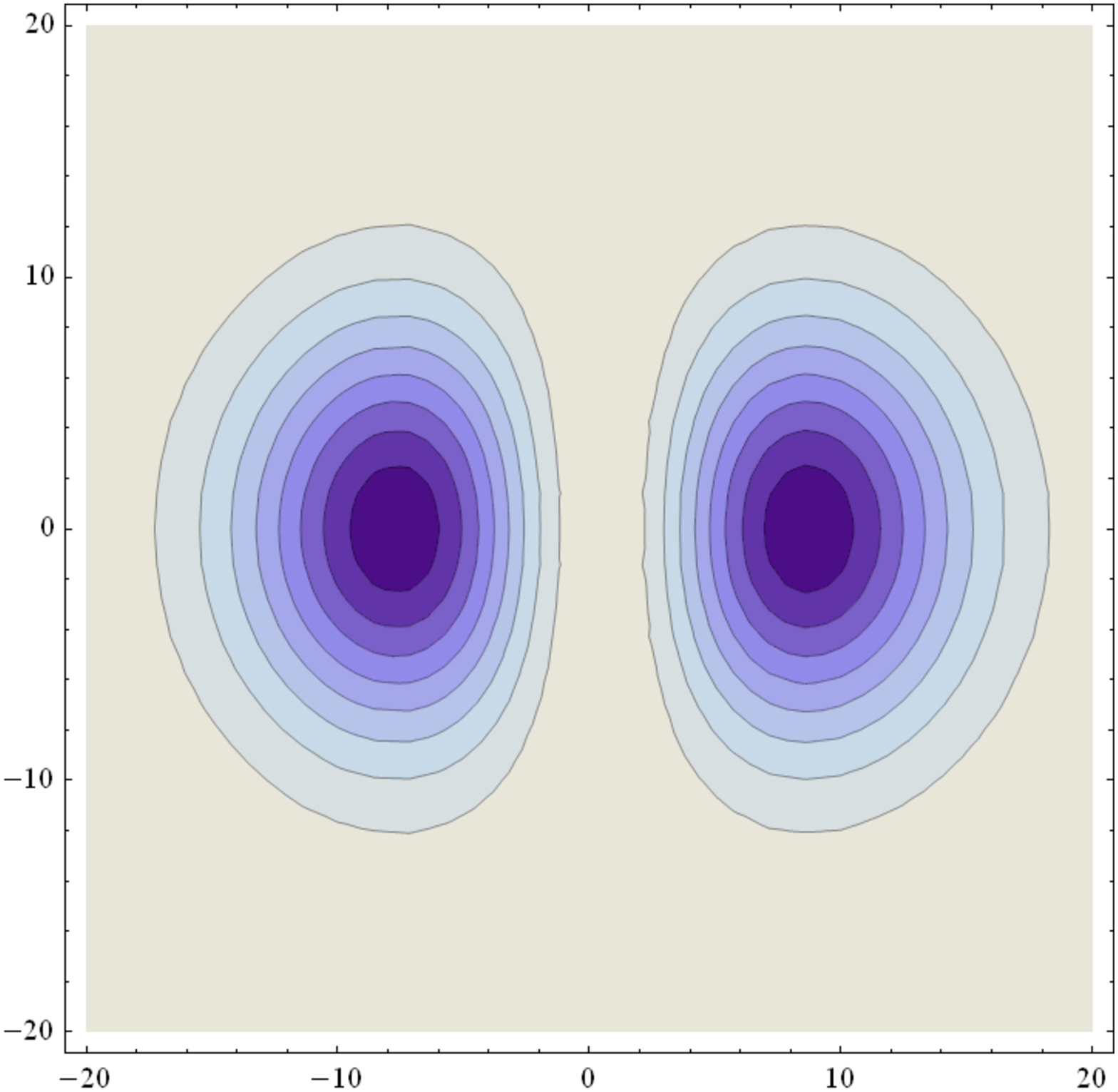}}
\caption{
The comparison of mode profiles of ghosts from the tensor model and 
the general relativity for $D=2$, through $\delta K_{ab}$ in (\ref{eq:tensordk})
and (\ref{eq:metricdk}).
The left couple of figures show the profiles of the modes 
in the lower and upper trajectories in the tensor model.
The right couple of figures show the profiles of the normal 
($p_\mu n^\mu=0$) and the longitudinal ($p_\mu\propto n_\mu$) modes
in the general relativity. }  
\label{fig:ghostprofile}
\end{figure}

The mode profiles have also been checked as shown in Figure \ref{fig:ghostprofile}
for $D=2$.
The comparison of ghosts 
between the numerical analysis of the tensor model and the general 
relativity has also been done for $D=1,3$, 
and clear agreement has been obtained.

\section{Summary and future problems}
In a series of papers, I have studied the tensor model with actions which possess
the Gaussian backgrounds as their classical solutions.
These backgrounds represent fuzzy flat spaces with arbitrary dimensions,
and the small fluctuations around them have been compared with the general 
relativity on flat backgrounds through numerical analyses.
The numerical analyses for dimensions $D=1,2,3,4$
have shown that the long-wavelength low-lying fluctuation spectra 
are in one-to-one correspondence
with the geometric fluctuations in the general relativity.
The analyses have also shown that part of the orthogonal symmetry of
the tensor model spontaneously broken by the backgrounds corresponds to the 
local gauge symmetry (local translational symmetry) of the general relativity.
These results should be valid in all dimensions, because of the 
dimensional independence of the framework and of the way of analysis. 
Thus, the tensor model provides an interesting model of simultaneous
emergence of space, the general relativity and its gauge symmetry of
translation in general dimensions.

There seem to exist various questions about the results. 
For example, the agreement between the modes in the tensor model 
and in the general relativity should be checked also for higher orders of fluctuations,
although such higher order agreement is expected, because the general relativity 
(possibly with modified actions)
is the only theory of symmetric rank-two tensor field with the gauge symmetry.  
This will require more efficient numerical facility and/or technical 
developments. 
Another important question is the range of generality of the results,
which have only been shown so far around the Gaussian 
backgrounds in a few actions of examples. 
For the tensor model to be really interesting, such emergence should be shown to 
be common phenomena in more general settings.

%





\end{document}